\documentstyle[12pt]{article}

\newcommand{\be}{\begin{equation}}
\newcommand{\ee}{\end{equation}}
\newcommand{\ds}{\displaystyle}

\begin{document}

\hfill {\large ITP--97--4E}

\vspace{3cm}

\begin{center}
{\Large \bf Polarization observables in $A(d,p)$ breakup
and quark degrees of freedom in the deuteron}\\[1.cm]
{\large A.P.Kobushkin \footnote{e-mail address: akob@ap3.gluk.apc.org}}\\
{\it Bogolyubov Institute for Theoretical Physics}\\
{\it 252143, Kiev, Ukraine}
\end{center}

\vspace{2cm}
\begin{center}
{\large Abstract}
\end{center}
The differential cross section, the tensor analyzing power,
$T_{20}$, and the polarization transfer, $\kappa_0$, in
$^{12}C(d,p)$ breakup at relativistic energy are calculated
within a model which incorporates multiple scattering
and Pauli principle at quark level. It is shown that the
rescattering and quark exchange affect drastically
the polarization observables for kinematical
region corresponding to high internal momentum in the deute\-ron.

\vspace{1cm}
\mbox{}

\newpage

During last 15 years detailed data on $A(d,p)$ differential
cross section (DCS) \cite{Ableev83}--\cite{Ableev92}, tensor analizing
power, $T_{20}$ \cite{Perdrisat}--\cite{Azhgirey.I}, and
coefficient of polariza\-tion transfer, $\kappa_0$
\cite{Nomofilov},\cite{Cheung}--\cite{Kuehn}, have been obtained.
These data together with data obtained from electromagnetic
probe, give important information about the deuteron structure
in a wide region, including that of a smooth transition from
nucleon--meson to quark--gluon picture.

Although from the very beginning it was evident that multiple
scattering (MS)
of the deuteron constituents on target nucleons should significantly
affect theoretical interpretation of $A(d,p)$ data \cite{Bertocchi},
there were brought up some arguments  that it would only
renormalize the results of impulse approximation (IA) without
strong modification of its shape \cite{Ableev83}, \cite{FS}, \cite{KV}.
Following this idea, the MS was often ``forgotten'' in different
theoretical models; for example, in
estimations of some reaction mechanisms beyond IA \cite{Lykasov},
model incorporation of quasi--free channels \cite{Mueller},
reduced matrix elements in QCD \cite{Kob92}, calculations within
the Bethe--Salpeter formalism \cite{Kaptari}, contribution of quark
exchange (QE) in the deuteron \cite{KSG}, etc. 
Direct calculations of MS effects in the DCS demonstrated that it
does not change the universal $A^{2/3}$ dependence of the DCS and
cannot explain experimental data without modification
of the deuteron wave function (DWF) at short distances \cite{Ableev92}.
From the other hand
it should be also mensioned that the recent
precise measurements of $T_{20}$ demonstrates that it has a systematic
difference for $^1H$ and $^{12}C$ targets, despite the fact that
qualitatively it has  similar behavior for different target
nuclei \cite{Azhgirey.I}.

Attempts to take into account MS effects \cite{PP}, as well as
those accom\-pained with a final state interaction \cite{Dakhno}, in
the DCS and $T_{20}$ in the $0^{\circ}$ inclusive
$^1H(d,p)$ breakup were done only in the framework of nonrelativistic
calculations. In the present paper we propose the first systematic
study of the MS based on relativistic calculations, together with
QE effects.

We start with the Bertocchi--Treleani approach \cite{Bertocchi} based
on the Sitenko--Glau\-ber MS theory. It includes elastic
rescattering of the deuteron consti\-tuents, the proton and neutron,
as well as inelastic collision of the constituent neutron.
We modify the Bertocchi--Treleani model in the following way:
\begin{itemize}
\item
the DFW is considered in the framework of ``minimal relativization
prescription'' with dynamics in the infinite momentum frame (IMF)
\cite{FS},\cite{KV};
\item
it takes into account the Pauli principle at the constituent quark level.
\end{itemize}

To specify the lab. frame we choose $y$-axis  along the quantization
one and $z$-axis
along the deuteron beam. In turn the IMF is defined as a
limiting reference frame moving, with respect to the lab. frame, in
the negative $z$-direction with velocity close to the speed of light.
In IMF the proton momentum is parametrized by the transversed
momentum, $\vec k_{\perp}\equiv (k_1,k_2) $ and the fraction of the
deuteron momentum carried by the proton in the longitudinal direction,
$\alpha$.  These momenta are expressed in the lab. frame as:
 $\vec k_{\perp}=\vec p_{\perp}$ and $\alpha=(E_p+p_3)/(E_d+d_3)$,
where $E_p$ and $E_d$ are lab. energy and $\vec p$ and $\vec d$ are
lab.  momenta of the proton and deuteron, respectively. In terms of
$\alpha$ and $\vec k_{\perp}$ the
invariant mass of virtual proton--neutron system is
\begin{equation}
M^2_{pn}=\frac{m^2_{\perp}}{\alpha(1-\alpha)}\neq m^2_d, \ \
 m^2_{\perp}=m^2_N+k^2_{\perp},
\label{1}
\end{equation}
where $m_d$ and $m_N$ are the deuteron and nucleon masses. In IMF
the argument of the relativistic DWF (usually called internal
momentum in the deuteron) is $\vec k=(\vec k_{\perp},k_3)$, where
\begin{equation}
k_3=\pm\sqrt{\varepsilon^2-m^2_{\perp}}=
m_{\perp}(\alpha-\frac{1}{2})/(\sqrt{\alpha(1-\alpha)}),\ \
\varepsilon=\frac{1}{2}M_{pn}
\label{2}
\end{equation}
and the relativistic DWF is
\begin{eqnarray}
&&\psi_{pn}^{Mmm^{\prime}}
=\sqrt{\frac{\epsilon}{m}}\chi^{\dag}_{m^{\prime}}
U^{\dag}(-{\vec k}){\cal M}^{(M)}U({\vec k})
\sigma_2 \chi_{m},
\label{3} \nonumber \\
&&{\cal M}^{(M)}=
\vec \epsilon^{\ (M)}\frac{i}{\sqrt{8\pi }} \left\{ {\vec \sigma}
u(k)+\sqrt{\frac{1}{2}}\left({\vec \sigma} -3\frac{\vec k
(\vec\sigma \vec k)}{k^2}\right)w(k) \right\} .
\label{4}
\end{eqnarray}
In (\ref{4}) $U(\vec k)$ is the matrix of relativistic spin rotation
(the Melosh transfor\-mation):
\begin{equation}
U({\vec k})=\frac{m+\epsilon+k_3+i({\vec\sigma \times \vec k})_3}
{\sqrt{(m+\epsilon+k_3)^2+k_{\perp}^2}},
\label{5}
\end{equation}
$\vec \epsilon^{\ (M)}$ is the nonrelativistic polarization vector
and $\chi_m$ and $\chi_{m^{\prime}}$ are Pauli spinors for the
proton and neutron, respectively.

The Pauli principle considered at the level of constituent quarks
modifies the DWF, which becomes equivalent to the
Resonating Group Method (RGM) wave function \cite{GBKK},\cite{GK}:
\begin{equation}
\psi^{d}(1,2,\ldots,6)=
\hat{A}\left\{\varphi_{N}(1,2,3)
\varphi_{N}(4,5,6)\chi ({\vec r})\right\},
\label{rgm}
\end{equation}
where $\hat{A}$ is the quark
antisymmetrizer and $\varphi_{N}$  are wave functions of the  nucleon
three quark ($3q$) clusters; $\chi ({\vec r})$ is the RGM distribution
function and  $\vec r$ stands for the relative coordinate between
the centers of masses of the  $3q$ bags.

Due to the presence of the antisymmetrizer in (\ref{rgm}) the DWF, being
de\-com\-posed into $3q \times 3q$ clusters, includes, apart from the
standard $pn$ component, nontrivial $NN_R$, $N_RN$
and $N_RN_{R^{\prime}}$ components which correspond to all possible
nucleon resonance states, $N_R$ (see \cite{GK}). Most of the this
isobars have negative parity and thus generate effective $P$ waves of the
deuteron \cite{KSG},\cite{GK}.

Following Refs.\cite{GK} we choose  $\chi(\vec r)$ as a conventional $NN$
DWF, $\chi_{NN}(\vec r)$, modified by the RGM
renormalization condition of \cite{WT}. Fig.1 displays the
renormalization of the $S$ wave for some deuteron wave functions;
the corresponding effect for the $D$ wave is less than 1\% \cite{KSG}.

The Pauli principle applied to the deuteron at the quark
level leads to the following modifications of the relativistic DWF
\begin{itemize}
\item
in the $d\to NN$ channel one has to change in (\ref{4})
$u(k)\to \tilde u(k)=u_{renorm.}(k)+\varphi(k)$;
\item
add wave functions of $d\to N_RN$ channels
$\psi_{N_Rp}^{Mmm^{\prime}}(\vec k)$,
which is the Fourier transformation of the overlap
\begin{equation}
\tilde{\psi}_{pN_R}({\bf r}_{p})
=\left(\frac{6!}{3!3!2}\right)^{1/2}
\langle\varphi_{N_R}(1,2,3)
\varphi_{p}(4,5,6)|\psi^{d}(1,2,\ldots,6)\rangle
\label{7}
\end{equation}
(with the appropriate spin projections $M$, $m$, $m^{\prime}$
on the quantization axis) multiplied by relativistic factor
$\sqrt{\varepsilon/m_N}$.
\end{itemize}
The explicit expressions for $\varphi(k)$ and some of
$\psi_{pN_R}({\vec k}_{p})$ can be found in Ref.\cite{KSG}.

With this modifications the Bertocchi--Treleani model for the
in\-va\-riant DCS of the $0^{\circ}$
inclusive $(d,p)$ breakup for pure spin states $M$ and $m$ of the
deuteron and proton, respectively, becomes:
\begin{eqnarray}
E_p\frac{d^3\sigma_M^m}{d\vec p}&\equiv&I^m_M=
\frac{C_d F(k)}{2(1-\alpha)^2}\sum_{N_{R}}\sum_{m^{\prime}}
\left\{
\sigma^T_{N_{R}A}\left|
\psi_{pN_{R}}^{Mmm^{\prime}}({\vec k}_{\perp}=0,k_3)\right|^2-
\right. \nonumber \\
&-&\left.2{\rm Re}\left(
\psi_{pN_{R}}^{\ast\ Mmm^{\prime}}({\vec k}_{\perp}=0,k_3)
\int d^2k_{\perp}
\psi_{pN_{R}}^{\ast\ Mmm^{\prime}}({\vec k})
\frac{d^2\tilde\sigma_{N_{R}A}}{d{\vec k}_{\perp}} \right)
+\right. \nonumber \\
&+&\left.
\int d^2 k_{\perp}\left|\psi_{pN_{R}}^{\ast\ Mmm^{\prime}}
({\vec k})\right|^2
\frac{d^2\sigma_{pA}}{d{\vec k}_{\perp}}
\right\},
\label{BT}
\end{eqnarray}
where summation over $N_{R}$ includes the
neutron and all resonance states gene\-rated by QE; $m^{\prime}$ is
the spin projection of the $N_{R}$, $C_d$ is the
renormalization coefficient introduced by Bertocchi and
Treleani \cite{Bertocchi}, $F(k)$ is the ratio of flux
factors for $nA$ and $dA$ collisions, $\sigma^T_{N_{R}A}$
is the total and $d^2\sigma_{N_{R}A}/d^2 {\vec k}_{\perp}$
and $d^2\sigma_{pA}/d^2 {\vec k}_{\perp}$ are differential cross
sections of the nucleus $A$, respectively.
In our calculations we use for $N_R$ the lowest 10 states with effective
numbers larger then $10^{-4}$.

The invariant DCS, $I$,
the tensor analyzing power, $T_{20}$, and the polari\-zation transfer
coefficient, $\kappa_0$, are
\begin{equation}
I=\frac{1}{3}\sum_{M,m}I^m_M, \ \
T_{20}=-\sqrt{2}
\frac{I_1^+ +I_1^- - 2I_0^+}{I_1^+ +I_1^- +I_0^+}, \ \
\kappa_0= \frac{I_1^+ +I_1^- }{I_1^+ +I_1^- +I_0^+}.
\label{kappa}
\end{equation}

In numerical calculations we take for the total cross section
$\sigma^T_{n^{12}C}$ the experimental value for
$\sigma^T_{p^{12}C}=340$ mb. The differential cross sections
 $d^2\tilde \sigma_{N^{12}C}/d{\vec k}_{\perp}\ $ and
$d^2 \sigma_{n^{12}C}/d\vec{k}_{\perp}\ $ were calculated in the
framework $\ $ of the Sitenko--Glauber model and them
approximated by universal
curve\\ $A_1 e^{-B_1k^2_{\perp}}+A_2 e^{-B_2k^2_{\perp}}$, with
$A_i=$2611.3 and 119.3 mb/(GeV/c)$^2$,
$B_i=$69.3 and 4.3 (GeV/c)$^{-2}$, respectively.
For $N_R$ we have used similar dependence
assuming the $\sigma^T_{N_R\ ^{12}C}$, $A_i^R$, and $B_i^R$
as  free parameters.

In Figs.2 and 3 we compare results of our calculations with
experimental data for $I$, $T_{20}$ and $\kappa_0$.
The parameters for average resonance were chosen as
$\sigma^T_{N_R\ ^{12}C}=400$ mb,
$A_1^R=2373.9$ mb/(GeV/c)$^2$, $B_1^R=77.0$ (GeV/c)$^{-2}$,
$A_2^R=A_2$ and $B_2^R=B_2$.
One concludes that QE and MS
give large contribution in the DCS and
polarization observables of the $0^{\circ}$ inclusive $^{12}C(d,p)$
breakup for the kinematical region corresponding to high internal
momentum $k$ and qualitatively reproduce the behavior of
this quantities. The Melosh transformation does not affects the final
results significantly.

Note that in our calculations we  ignore  $NA\to NA$ spin--flip
amplitude. There are some arguments that $np$ spin--flip is important
for $T_{20}$ in $^1H(d,p)$ \cite{PP}. This, as well as detailed
$A$-dependence of the polarization observables in $A(d,)p$, remains
an open question to the theory and we hope to discuss this further in a
separate paper.\\

The author would like to thank E.A.~Strokovsky for numerous discussions
at every stage of this investigation, as well as
C.F.~Perdrisat and V.~Punjabi for reading maniscript and useful
comments. He is also indebted for helpful conver\-sa\-tions with
L.G.~Dakhno that strongly stimulated the work.

\newpage
\includegraphics{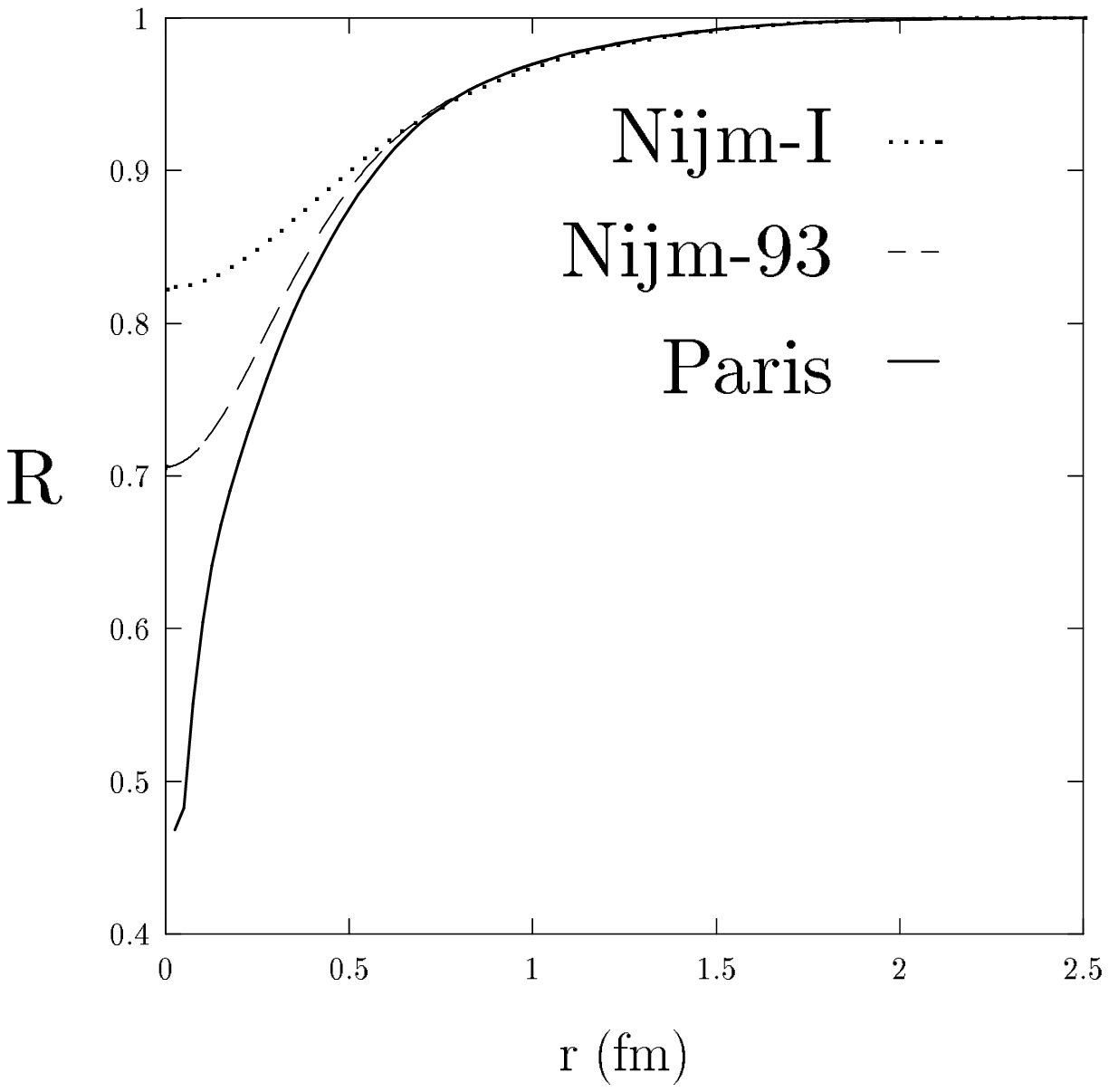}
\mbox{}
\vfill \begin{center}{\LARGE Fig.~1.}\end{center}

\newpage
\includegraphics{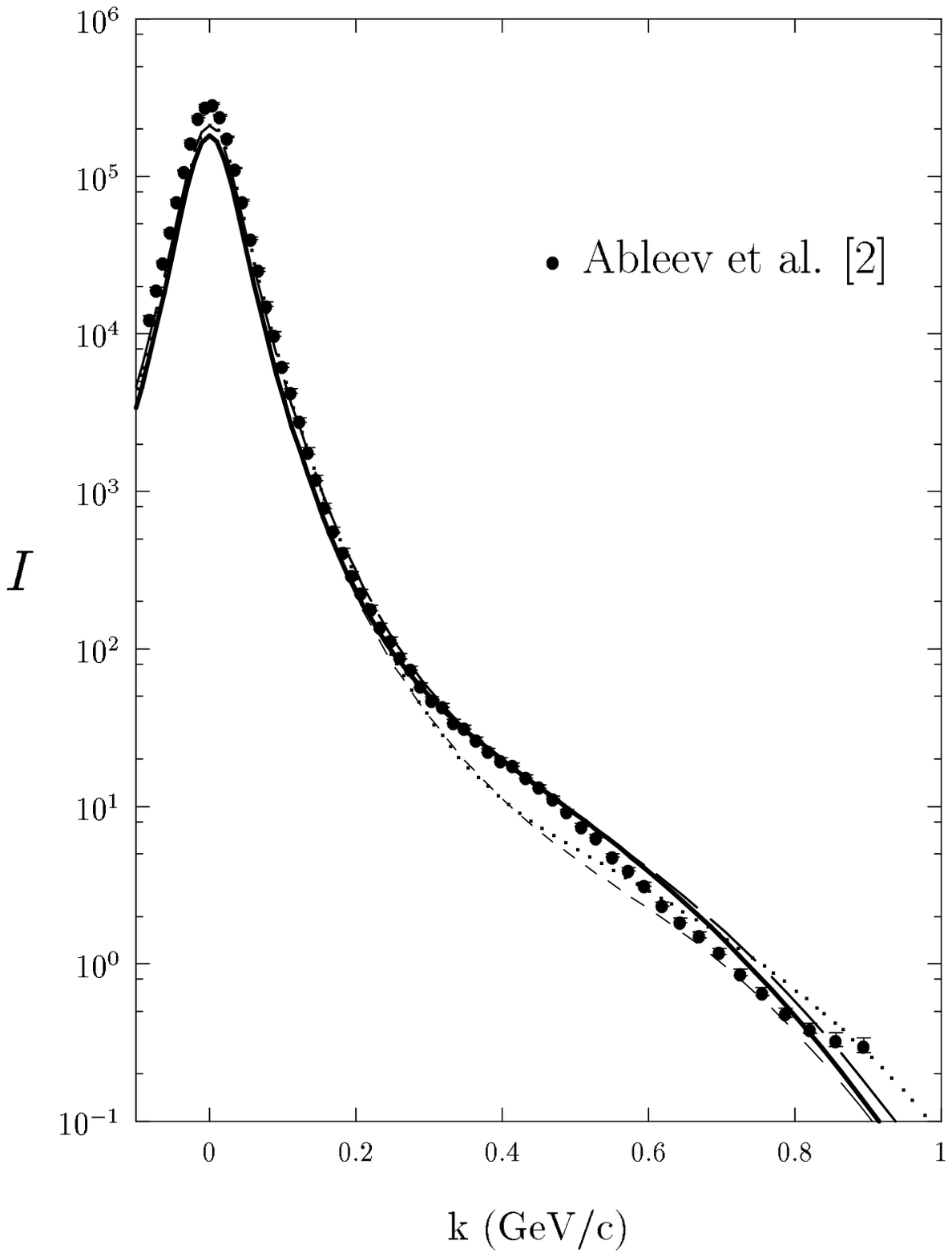}
\mbox{}
\vfill \begin{center}{\LARGE Fig.~2.}\end{center}

\newpage
\includegraphics{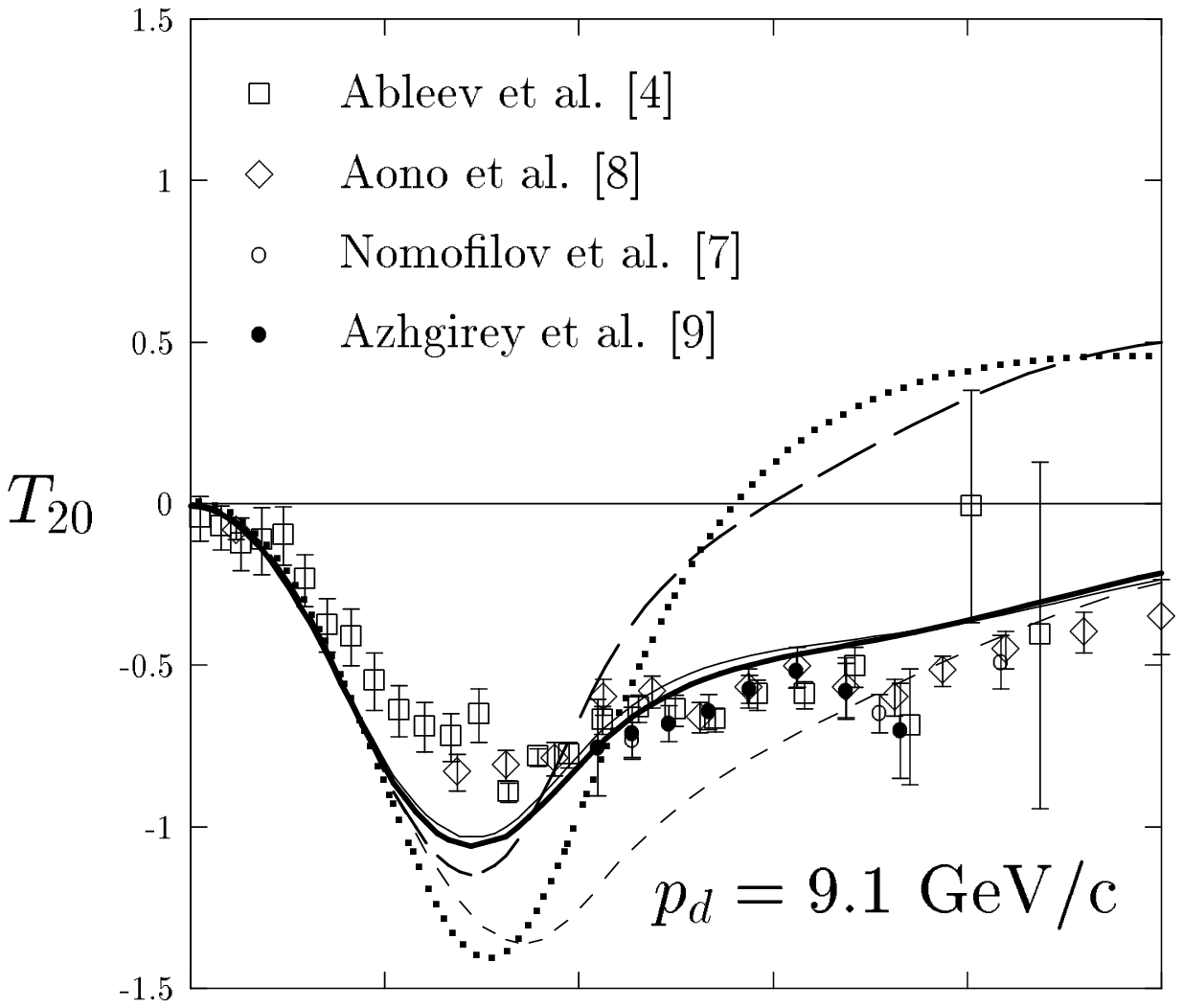}

\includegraphics{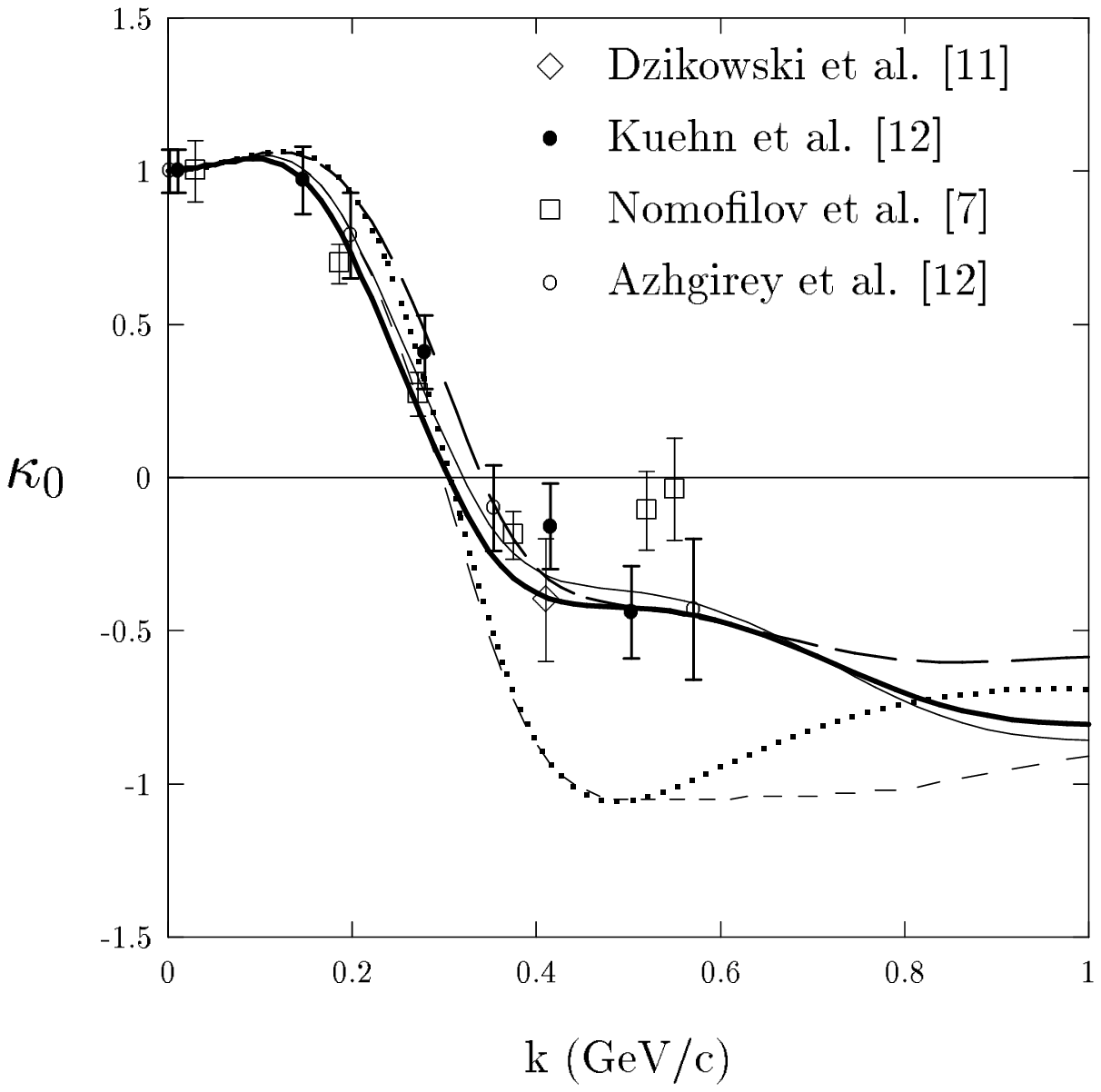}
\mbox{}
\vfill \begin{center}{\LARGE Fig.~3.}\end{center}

\newpage
\begin{center}
\Large Figure captions
\end{center}

\mbox{}\\
 Fig. 1. Ration $R=u_{renorm.}(r)/u(r)$ for some deuteron
 wave functions (PARIS --- \cite{Paris}, Nijm-93 and
Nijm-I --- \cite{deSwart} and
see at URL address\\
http://NN-online.sci.kun.nl/NN/deut.html)
for radius of quark core $b=0.8$ fm.
\\[0.5cm]
 Fig. 2. The invariant differential cross section (DCS)
 $I\equiv E_p\ds\frac{d^3\sigma}{d\vec p}$ of the $0^{\circ}$
 inclusive $^{12}C(d,p)$ breakup at $p_d=$9.1 GeV/c plotted
 versus the relativistic internal momentum in the deuteron
 $k$ (see text for definition). Curves show results of
calculations with the Nijm-I deuteron wave function in the framework
of multiple scattering: with  (bold solid line) and without
quark exchange (short-dashed line); and  IA: with (long dashed line)
and without quark exchange (dotted line). Contribution of the Melosh
 transformation in DCS is negligible and not shown. DCS is given in
mb$\times$GeV/(GeV/c)$^3$/srad.
\\[0.5cm]
 Fig. 3. The tensor analyzing power $T_{20}$ (upper panel)
 and the polarization transfer $\kappa_0$ (down panel) in the $0^{\circ}$
 inclusive $^{12}C(d,p)$ breakup as function of the relativistic
 internal momentum in the deuteron $k$. Curves are the same as in fig.~2,
thin solid lines are the result of multiple scattering with quark exchange
omitting relativistic spin rotation ($U(\vec k)=1$).

\end{document}